\documentclass[pmlr]{jmlr}

\usepackage{longtable}

\usepackage{array}
\usepackage{multirow}

\usepackage{booktabs}
\usepackage[load-configurations=version-1]{siunitx} 

\makeatletter
\def\set@curr@file#1{\def\@curr@file{#1}} 
\makeatother

\theorembodyfont{\upshape}
\theoremheaderfont{\scshape}
\theorempostheader{:}
\theoremsep{\newline}

\title[Identifying Radiological Findings]{Identifying Radiological Findings Related to COVID-19 from Medical Literature}

\author{\Name{Yuxiao Liang}
       \Email{yul154@eng.ucsd.edu}\\ 
       \addr UC San Diego\\
       \AND
       \Name{Pengtao Xie}
       \Email{pengtaoxie2008@gmail.com}\\ 
       \addr UC San Diego\\
       }

\begin{document}

\maketitle

\begin{abstract}
Coronavirus disease 2019 (COVID-19) has infected more than one million individuals all over the world and caused more than 55,000 deaths, as of April 3 in 2020. Radiological findings are important sources of information in guiding the diagnosis and treatment of COVID-19. However, the existing studies on how radiological findings are correlated with COVID-19 are conducted separately by different hospitals, which may be inconsistent or even conflicting due to population bias. To address this problem, we develop natural language processing methods to analyze a large collection of COVID-19 literature containing study reports from hospitals all over the world, reconcile these results, and draw unbiased and universally-sensible conclusions about the correlation between radiological
findings and COVID-19. We apply our method to the CORD-19 dataset and successfully extract a set of radiological findings that are closely tied to COVID-19.
\end{abstract}

\section{Introduction}
Coronavirus disease 2019 (COVID-19) is an infectious disease that has affected more than one million individuals all over the world and caused more than 55,000 deaths, as of April 3 in 2020. The science community has been working very actively to understand this new disease and make diagnosis and treatment guidelines based on the findings. One major stream of efforts are focused on discovering the correlation between radiological findings in the lung areas and  COVID-19. There have been several works~\citep{liu2020clinical,li2020coronavirus} publishing such results. However, existing studies are mostly conducted separately by different hospitals and medical institutes. Due to geographic affinity, the populations served by different hospitals have different genetic, social, and ethnic characteristics. As a result, the radiological findings from  COVID-19 patient cases in different populations are different. This population bias incurs inconsistent or even conflicting conclusions regarding the correlation between radiological findings and  COVID-19. As a result, medical professionals cannot make informed decisions on how to use radiological findings to guide diagnosis and treatment of  COVID-19.

We aim to address this issue. Our research goal is to develop natural language processing methods to collectively analyze the study results reported by many hospitals and medical institutes all over the world, reconcile these results, and make a holistic and unbiased conclusion regarding the correlation between radiological findings and COVID-19. Specifically, we take the CORD-19 dataset~\citep{cord19}, which contains over 45,000 scholarly articles, including over 33,000 with full text, about COVID-19, SARS-CoV-2, and related coronaviruses. We develop sentence classification methods to identify all sentences narrating radiological findings from COVID-19. Then constituent parsing is utilized to identify all noun phrases from these sentences and these noun phrases contain abnormalities, lesions, diseases identified by radiology imaging such as X-ray and computed tomography (CT). We calculate the frequency of these noun phrases and select those with top frequencies for medical professionals to further investigate. Since these  clinical entities are aggregated from a number of hospitals all over the world, the population bias is largely mitigated and the conclusions are more objective and universally informative. From the CORD-19 dataset, our method successfully discovers a set of clinical findings that are closely related with COVID-19.

The major contributions of this paper include:
\begin{itemize}
    \item We develop natural language processing methods to perform unbiased study of the correlation between radiological findings and COVID-19. 
    \item We develop a bootstrapping approach to effectively train a sentence classifier with light-weight manual annotation effort. The sentence classifier is used to extract radiological findings from a vast amount of literature.
    \item We conduct experiments to verify the effectiveness of our method. From the CORD-19 dataset, our method successfully discovers a set of clinical findings that are closely related with COVID-19.
\end{itemize}
The rest of the paper is organized as follows. In Section 2, we introduce the data. Section 3 presents the method. Section 4 gives experimental results. Section 5 concludes the paper.

\section{Dataset}
We used the COVID-19 Open Research Dataset (CORD-19)~\citep{cord19} for our study. In response to the COVID-19 pandemic, the White House and a coalition of research groups  prepared the CORD-19 dataset. It contains over 45,000 scholarly articles, including over 33,000 with full text, about COVID-19, SARS-CoV-2, and related coronaviruses.  These articles are contributed by hospitals and medical institutes all over the world. Since the outbreak of COVID-19 is after November 2019, we select articles published after November 2019 to study, which include a total of 2081 articles and about 360000 sentences.
Many articles report the radiological findings related to COVID-19. Table~\ref{dataset_example} shows some examples. 

\begin{table}[htbp]
  \centering 
  \caption{Example sentences describing radiological findings related with COVID-19}
  \begin{tabular}{|m{14cm}|}\hline
  \small The X-ray chest and Highresolution computed-tomography chest showed a progressively reduced bilateral pleural effusion, interstitial-alveolar edema with bilateral hilar congestion. \\ \hline
  \small Non-pleural effusion appeared in the first and second CT examination and a bilateral pleural effusion appeared in the third CTexamination. \\ \hline
  \small Chest CT scans revealed that the grade 2 lesion and one of the grade 3 lesions as areas of prominent breast tissue mimicking faint GGO and post-inflammatory focal atelectasis on chest radiography, respectively. \\ \hline
  \small A showed many CT features such as ground-glass opacification,cobblestone/reticular pattern(blue arrow and bule line around region), frbrosis and dilated bronchi with thickened wall (the enlarged image in the upper right corner, red arrow). \\ \hline
  \small The typical CT images show bilateral pulmonary parenchymal ground-glass and consolidative pulmonary opacities, sometimes with a rounded morphology and a peripheral lung distribution. \\ \hline
  \small The chest CT ( Figure 3 ) showed multi-focal GGO with parenchyma consolidation and subpleural effusion, predominantly involving upper lungs.\\ \hline
  
  \end{tabular}
  \label{dataset_example} 
\end{table}


\section{Methods}
Our goal is to develop natural language processing (NLP) methods to analyze a large collection of COVID-19 literature and discover unbiased and universally informative  correlation between radiological findings and COVID-19. To achieve this goal, we need to address two technical challenges. First, in the large collection of COVID-19 literature, only a small part of sentences are about radiological findings. It is time-consuming to manually identify these sentences. Simple methods such as keyword-based retrieval will falsely retrieve sentences that are not about radiological findings and miss sentences that are about radiological findings. How can we develop NLP methods to precisely and comprehensively extract sentences containing radiological findings with minimum human annotation? Second, given the extracted sentences, they are still highly unstructured, which are difficult for medical professionals to digest and index. How can we further process these sentences into structured information that is more concise and easy to use?

To address the first challenge, we develop a sentence classifier to judge whether a sentence contains radiological findings. To minimize manual-labeling overhead, we propose easy ways of constructing positive and negative training examples, develop a bootstrapping approach to mine hard examples, and use hard examples to re-train the classifier for reducing false positives. To address the second challenge, we use constituent parsing to recognize noun phrases which contain critical medical information (e.g., lesions, abnormalities, diseases) and are easy to index and digest. We select noun phrases with top frequencies for medical professionals to further investigate.



\subsection{Extracting Sentences Containing Radiological Findings}
In this section, we develop a sentence-level classifier to determine whether a sentence contains radiological findings. To build such a classifier, we need to create positive and negative training sentences, without labor-intensive annotations. To obtain  positive examples, we resort to the MedPix\footnote{https://medpix.nlm.nih.gov/home} database, which contains radiology reports narrating radiological findings. MedPix is an open-access online database of medical images, teaching cases, and clinical topics. It contains more than 9000 topics, 59000 images from 12000 patient cases. We selected diagnostic reports for CT images and used sentences in the reports as  positive samples. To obtain negative sentences, we randomly sample some sentences from the articles and quickly screen them to ensure that they are not about radiological findings. Since most sentences in the literature are not about radiological findings, a random sampling can almost ensure the select sentences are negative. A manual screening is conducted to further ensure this and the screening effort is not heavy.

\begin{figure}[htbp]
  \centering 
  \includegraphics[width=3in]{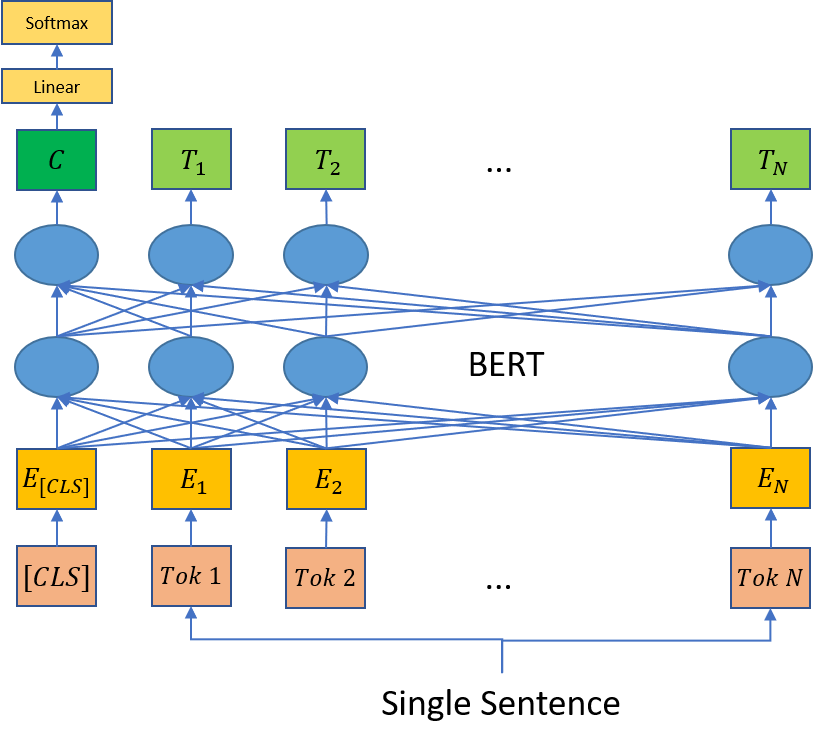} 
  \caption{Architecture of the sentence classification model.}
  \label{BERT} 
\end{figure}

Given these positive and negative training sentences, we use them to train a sentence classifier which predicts whether a sentence is about the radiological findings of COVID-19. We use the Bidirectional Encoder Representations from Transformers (BERT) \citep{devlin2018bert} model for sentence classification. BERT is a neural language model that learns contextual representations of words and sentences. BERT pretrains deep bidirectional representations from
unlabeled text by jointly conditioning on both
left and right context in all layers. To apply the pretrained BERT to a downstream task such as sentence classification, one can add an additional layer on top of the BERT architecture and train this newly-added layer using the labeled data in the target task. In our case, similar to \citep{lee2020biobert}, 
we pretrain the BERT model on a vast amount of biomedical literature to obtain semantic representations of words. A linear layer is added to the output of BERT for predicting whether this sentence is positive (containing radiological finding) or negative. The architecture and hyperparameters of the BERT model used in our method are the same as those in \citep{lee2020biobert}. Figure~\ref{BERT} shows the architecture of the classification model.




When applying this trained sentence classifier to unseen sentences, we found that it yields a lot of false positives: many sentences irrelevant to radiological findings of COVID-19 are predicted as being relevant. To solve this problem, we iteratively perform hard example mining in a bootstrapping way and use these hard examples to retrain the classifier for reducing false positives. At iteration $t$, given the classifier $C_t$, we apply it to make predictions on unseen sentences. Each sentence is associated with a prediction score where a larger score indicates that this sentence is more likely to be positive. We rank these sentences in descending order of their prediction scores. Then for the top-K sentences with the largest prediction scores, we read them and label each of them as either being positive or negative. Then we add the labeled pairs to the training set and re-train the classifier and get $C_{t+1}$. This procedure is repeated again to identify new false positives and update the classifier using the new false positives.



\subsection{Extracting Noun Phrases}

\begin{figure}[htbp]
  \centering 
  \includegraphics[width=4 in]{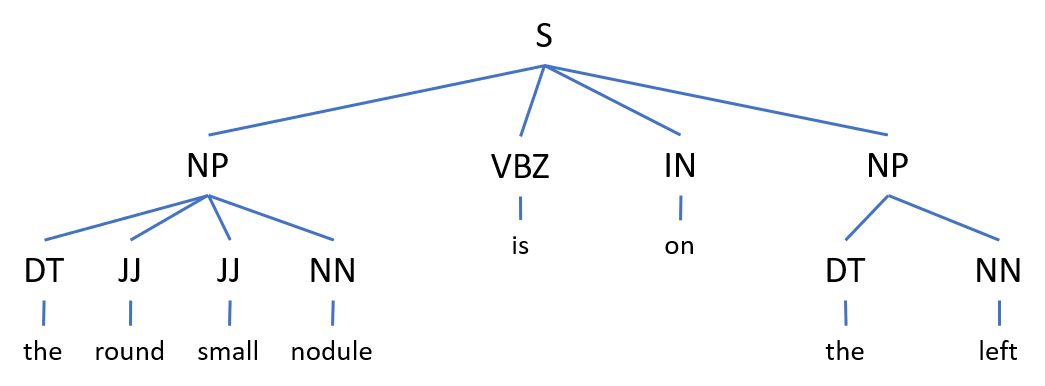} 
  \caption{An example of constituent parse tree.}
  \label{chunker} 
\end{figure}

The extracted sentences containing radiological findings of COVID-19 are highly unstructured, which are still difficult to digest for medical professionals. To solve this problem, from these unstructured sentences, we extract structured information that is both clinically important and easy to use. We notice that important information, such as lesions, abnormalities, diseases, is mostly contained in noun phrases. Therefore, we use NLP to extract noun phrases and perform further analysis therefrom. First, we perform part-of-speech (POS) tagging to label each word in a sentence as being a noun, verb, adjective, etc. Then on top of these words and their POS tags, we perform constituent  parsing to obtain the syntax tree of the sentence. An example is shown in Figure~\ref{chunker}. From bottom to top of the tree, fine-grained linguistic units such as words are composed into coarse-grained units such as phrases, including noun phrases. We obtain the noun phrases by reading the node labels in the tree. 

Given the extracted noun phrases, we remove stop words in them and perform lemmatization to eliminate non-essential linguistic variations. We count the frequency of each noun phrase and rank them in descending frequency. Then we select the noun phrases with top frequencies and present them to medical professionals for further investigation.



\section{Experiment}

\subsection{Experimental Settings} 

For building the initial sentence classifier (before hard-example mining), we collected 2350 positive samples from MedPix and 3000 negative samples from CORD-19.  We used 90\% sentences for training and the rest 10\% sentences for validation. The weights in the sentence classifier are optimized using the Adam~\cite{} algorithm with a learning rate of $2\times 10^{-5}$ and a mini-batch size of 4. In bootstrapping for hard example mining, we added 400 false positives in each iteration for classifier-retraining and we performed 4 iterations of bootstrapping.

\subsection{Results of Sentence Classification} 

Under the final classifier, 998 sentences are predicted as being positive. Among them, 717 are true positives (according to manual check). The classifier achieves a precision of 71.8\%. For the initial classifier (before adding mined hard examples using bootstrapping), among the top 100 sentences with the largest prediction scores, 53 are false positives. The initial classifier only achieves a precision of 47\%. The precision achieved by classifiers trained after round 1-3 in bootstrapping is 55\%, 57\%, and 69\% respectively, as shown in Table~\ref{precision_of_top100}. This demonstrates the effectiveness of hard example mining.  Table~\ref{cls_result_example} shows some example sentences that are true positives, true negatives, and false positives, under the predictions made by the  final classifier.

\begin{table}[htbp]
  \centering 
  \caption{Precision at top-100 after each iteration}
  \begin{tabular}{|c|c|}\hline
    Iteration & Precision \\ \hline
    0 & 47\% \\ \hline
    1 & 55\% \\ \hline 
    2 & 57\% \\ \hline
    3 & 69\% \\ \hline
  \end{tabular}
  \label{precision_of_top100} 
\end{table}

\begin{table}[htbp]
  \centering 
  \caption{Example sentences for true positive (TP), true negative (TN), and false positive (FP).}
  \begin{tabular}{|p{0.9cm}|p{13cm}|}\hline
    \multirow{4}*{\textbf{ TP}} & \small Her chest radiograph demonstrated no abnormalities, but a CT scan of her chest revealed bilateral multifocal infiltrates and mediastinal and hilar lymphadenopathy. \\ \cline{2-2}
    ~ & \small The predominant pattern was groundglass opacity, with illdefined margins, air bronchograms, smooth or irregular inter lobular or septal thickening, and thickening of the adjacent pleura. \\ \cline{2-2}
    ~ & \small The patient's chest CT showed pulmonary consolidation, interlobular septal thickening, and pleural effusion. \\ \cline{2-2}
    ~ & \small Dynamic imaging showed progressive multi-spot patchy shadows in both lungs. \\ \hline
    \multirow{4}*{\textbf{ TN}} &
    \small Fourth, leisure activities and training on how to relax were properly arranged to help staff reduce stress. \\ \cline{2-2}
    ~ & \small Uncertainties of discrete E. coli samples and flow measurements were N30 and 97\%, respectively. \\ \cline{2-2}
    ~ & \small Acute lower respiratory tract infections (ALRIs) are a common illness in children $<$ 5 years old, with significant morbidity and mortality in infants and young children under the age of two (1) (2) (3) \\ \cline{2-2}
    ~ & \small Bovine viral diarrhea (BVD), which is caused by BVD virus (BVDV) infection, is one of the most important viral diseases of cattle, causing enormous economic losses to the livestock industry worldwide (Suda et al., 2018) . \\ \hline
    \multirow{4}*{\textbf{ FP}} &
    \small Pathology: At necropsy, there is diffuse lymphoma affecting multiple organs that can include those noted above plus kidney, adrenal gland, tonsils, and lung. \\ \cline{2-2}
    ~ & \small The stoma site per se is an incisional hernia by definition as it is a defect in the abdominal fascia with protruding intra-abdominal contents. \\ \cline{2-2}
    ~ & \small He was diagnosed with coronary artery disease in last year and was stented in Left anterior descending artery. \\ \cline{2-2}
    ~ & \small Ultimately, the patient underwent mitral valve replacement following the stress test. \\ \hline
    
  \end{tabular}
  \label{cls_result_example} 
\end{table}




\subsection{Results of Noun Phrase Extraction}
Table~\ref{NP} shows the extracted noun phrases with top frequencies that are relevant to radiology. Medical professionals can look at this table and select noun phrases indicating radiological findings for further investigation, such as consolidation, pleural effusion, ground glass opacity, thickening, etc. We mark such noun phrases with bold font in the table. To further investigate how a noun phrase is relevant to COVID-19, medical professionals can review the sentences mentioning this noun phrase. Table~\ref{noun_phrases_sentences_example_0},\ref{noun_phrases_sentences_example_1},\ref{noun_phrases_sentences_example_2} show some examples. 
\begin{table}[htbp]
  \centering 
  \caption{Extracted radiology-relevant noun phrases with top phrases. Noun phrases relevant to clinical findings are highlighted with bold font.}
  \begin{tabular}{|l|c|l|c|}\hline
  \textbf{Noun phrase} & \textbf{Frequency }& \textbf{Noun phrase} & \textbf{Frequency} \\ \hline
  ct & 103 & liver & 19 \\ \hline
  \textbf{consolidation} & 79 & \textbf{fibrosis} & 13 \\ \hline
  lung & 62 & \textbf{abscess} & 13 \\ \hline
  lobe & 61 & infection & 13 \\ \hline
  chest & 57 & \textbf{air bronchogram} & 12 \\ \hline
   \textbf{pleural effusion} & 54 & \textbf{lymphadenopathy} & 12 \\ \hline
   \textbf{ground glass opacity} & 53 & hrct & 11 \\ \hline
  lesion & 50 & interlobular septa & 11 \\ \hline
  chest x-ray & 38 & cxr & 10 \\ \hline
  pneumonia & 35 & trachea & 10 \\ \hline
  chest radiographs & 29 & \textbf{bronchiectasis} & 9 \\ \hline
  tomography & 27 & fever & 5 \\ \hline
  echocardiogram & 22 &\textbf{multiple sclerosis} & 3 \\ \hline
   \textbf{thickening} & 21 & & \\ \hline
  \end{tabular}
  \label{NP} 
\end{table}
\begin{table}[htbp]
\centering
  \caption{Example sentences containing interested noun phrases}
  \begin{tabular}{|p{3cm}|p{12cm}|}\hline
  \multirow{5}*{\begin{minipage}{3cm}\textbf{consolidation}\end{minipage}} & \footnotesize Chest CT images showed diffuse irregular small diffuse ground-glass nodular opacities with partial \textbf{consolidation} in bilateral lungs on Day 10 ( Figure 1A. \\ \cline{2-2}
    ~ & \footnotesize Axial thin-section non-contrast CT scan shows diffuse bilateral confluent and patchy ground-glass, air-bronchogram and \textbf{consolidation}, characterized by peripheral distribution. \\ \cline{2-2}
    ~ & \footnotesize B showed \textbf{consolidation} (pink arrow) and mixed pattern (yellow arrow and yellow line around region) in the bilateral lower lobes. \\ \cline{2-2}
    ~ & \footnotesize Subsequent chest CT images showed bilateral groundglass opacity, whereas the \textbf{consolidation} was resolved. \\ \cline{2-2}
    ~ & \footnotesize CT scan revealed bilateral peribronchial \textbf{consolidation}, swollen jejunum lymph node with uniform distribution of contrast medium, and multiple prominent nodules of the liver. \\ \hline
    
    \multirow{5}*{\begin{minipage}{3cm}\textbf{pleural effusion}\end{minipage}} & \footnotesize The chest film is not helpful in making a specific etiologic diagnosis; however, lobar consolidation, cavitation, and large \textbf{pleural effusions} support a bacterial cause. \\ \cline{2-2}
    ~ & \footnotesize Inhalation anthrax is notable for its absence of pulmonary infiltrate on chest imaging, but the presence of extensive mediastinal lymphadenopathy, \textbf{pleural effusions}, and severe shortness of breath, toxemia, and sense of impending doom. \\ \cline{2-2}
    ~ & \footnotesize An echocardiogram showed severe reduction of the left ventricular (LV) ejection fraction (25\%) and a chest X-ray exhibited bilateral \textbf{pleural effusion} with pulmonary interstitial edema (Fig. \\ \cline{2-2}
    ~ & \footnotesize The patient received 2 g/kg IVIG and significant \textbf{pleural effusion} was noted on the chest radiography and computed tomography 48 h after the completion of therapy. \\ \cline{2-2}
    ~ & \footnotesize Chest computed tomography showed bilateral \textbf{pleural effusion}, bilateral consolidation with air bronchogram, and ground-glass opacities in her left lung (Fig. \\ \hline
    
    \multirow{5}*{\begin{minipage}{3cm}\textbf{ground-glass opacity}\end{minipage}} & \footnotesize Axial thin-section non-contrast CT image shows diffusion lesions in bilateral lung, mainly manifested as \textbf{ground-glass opacification} (red arrow and red line around region),and cobblestone/Reticular pattern( blue arrow and blue line around region). \\ \cline{2-2}
    ~ & \footnotesize Chest CT will show areas of \textbf{ground-glass opacity} and consolidation in involved segments. \\ \cline{2-2}
    ~ & \footnotesize CT findings included ground glass opacity, consolidation, air bronchogram and nodular opacities. \\ \cline{2-2}
    ~ & \footnotesize CT images were reviewed and scored for lesion distribution, lobe and segment involvement, \textbf{ground-glass opacities}, consolidation, and interstitial thickening. \\ \cline{2-2}
    ~ & \footnotesize The most frequent radiological patterns on plain chest radiography was airspace/\textbf{ground glass opacification} in various lobes. \\ \hline
    
    \multirow{5}*{\begin{minipage}{3cm}\textbf{thickening}\end{minipage}} & \footnotesize Most of the lesions were distributed along the bronchovascular bundle or the dorsolateral and subpleural part of the lungs and were seen with or without interlobular septal \textbf{thickening}.  \\ \cline{2-2}
    ~ & \footnotesize CT images were reviewed and scored for lesion distribution, lobe and segment involvement, ground-glass opacities, consolidation, and interstitial \textbf{thickening}.  \\ \cline{2-2}
    ~ & \footnotesize Chest radiographs often demonstrate peribronchial \textbf{thickening} and infiltrates, often with areas of atelectasis. \\ \cline{2-2}
    ~ & \footnotesize Other c-HRCT signs of bronchiectasis include abnormalities in the surrounding lung may include parenchyma loss, emphysema, scars and nodular foci, 150 a linear array or cluster of cysts, dilated bronchi in the periphery of the lung, and bronchial wall \textbf{thickening} (Box 26.1). \\ \cline{2-2}
    ~ & \footnotesize CT showed multiple peripheral lesions in bilateral lungs, mainly characterized by GGO, paving stones and vascular \textbf{thickening}. \\ \hline

  \end{tabular}
  \label{noun_phrases_sentences_example_0} 
\end{table}

\begin{table}[htbp]

  \caption{Example sentences containing interested noun phrases}
  \begin{tabular}{|p{3cm}|p{11cm}|}\hline

    \multirow{5}*{\begin{minipage}{3cm}\textbf{fibrosis}\end{minipage}} & \footnotesize (D) Images obtained 5 days later showed partial absorption of the consolidation lesions in the right lower lobe, but \textbf{fibrosis}, bronchiectasis, and vascular thickening occurred. \\ \cline{2-2}
    ~ & \footnotesize High-resolution chest computer tomography (HRCT) presented diffuse ground glass opacities, intra-lobular reticulation and small cysts in the upper lobes, middle lobe and lingula, suggesting pulmonary \textbf{fibrosis} (Fig. \\ \cline{2-2}
    ~ & \footnotesize These findings indicated the appearance of interstitial changes, suggesting the development of \textbf{fibrosis}. \\ \cline{2-2}
    ~ & \footnotesize No other clinical or radiological changes of lung congestion, \textbf{fibrosis}, or cancer to explain these ground-glass lung changes, or any concomitant radiological changes of dense consolidation, pleural effusion, lymphadenopathy, or pneumomediastinum were seen. \\ \cline{2-2}
    ~ & \footnotesize If the diagnosis of idiopathic pulmonary \textbf{fibrosis} is not previously established this criterion can be met by the presence of radiologic and/or histopathologic changes consistent with usual interstitial pneumonia pattern on the current evaluation. \\ \hline
    
    \multirow{5}*{\begin{minipage}{3cm}\textbf{abscess}\end{minipage}} & \footnotesize A contrast-enhanced computed tomography scan is useful because it differentiates a fully developed \textbf{abscess} from cellulitis and delineates the full extent of the abscess.\\ \cline{2-2}
    ~ & \footnotesize The \textbf{abscess} pushes the adjacent tonsil downward and medially, and the uvula may be so edematous, as to resemble a white grape. \\ \cline{2-2}
    ~ & \footnotesize A CT of the neck was obtained to rule out other etiologies for respiratory distress, which was negative for peritonsillar or retropharyngeal \textbf{abscess}. \\ \cline{2-2}
    ~ & \footnotesize Abnormalities include bronchopneumonia, solitary or multiple lung nodes, miliary interstitial lung disease, lung \textbf{abscess}, and pleural effusion. \\ \cline{2-2}
    ~ & \footnotesize A lung abscess, like an \textbf{abscess} elsewhere, represents a localized collection of pus. \\ \hline
    
    \multirow{5}*{\begin{minipage}{3cm}\footnotesize\textbf{air bronchogram}\end{minipage}} & \footnotesize CT findings included ground glass opacity, consolidation, \textbf{air bronchogram} and nodular opacities. \\ \cline{2-2}
    ~ & \footnotesize Chest CT on Jan 13 showed improved status (3B) with diffuse consolidation of both lungs, uneven density and \textbf{air bronchogram}. \\ \cline{2-2}
    ~ & \footnotesize Chest computed tomography showed bilateral pleural effusion, bilateral consolidation with \textbf{air bronchogram}, and ground-glass opacities in her left lung (Fig. \\ \cline{2-2}
    ~ & \footnotesize The predominant pattern was groundglass opacity, with illdefined margins, \textbf{air bronchograms}, smooth or irregular inter lobular or septal thickening, and thickening of the adjacent pleura. \\ \cline{2-2}
    ~ & \footnotesize Computed tomography axial (A) and coronal (B) plane revealed multiple lesions (arrows) of ground glass opacity accompanied with consolidation under or near the pleura in bilateral lower lobes, with \textbf{air bronchogram} and thickened interlobular septa. \\ \hline
    
    \multirow{5}*{\begin{minipage}{3cm}\footnotesize\textbf{lymphadenopathy}\end{minipage}} & \footnotesize Merely consolidation, central distribution only, pleural effusions or \textbf{lymphadenopathy} were relatively rarely seen. \\ \cline{2-2}
    ~ & \footnotesize Thoracic and abdominal radiographs may provide evidence of pulmonary lesions, \textbf{lymphadenopathy}, and/or hepatosplenomegaly. \\ \cline{2-2}
    ~ & \footnotesize Neither pleural effusion nor \textbf{lymphadenopathy} was found. \\ \cline{2-2}
    ~ & \footnotesize Neither patient showed pleural effusion or \textbf{lymphadenopathy}. \\ \cline{2-2}
    ~ & \footnotesize 3.1 presence of \textbf{lymphadenopathy} (defined as lymph node size of !10 mm in short-axis dimension); 3.2 presence of pericardial effusion; 3.3 ascending thoracic aorta diameter. \\ \hline

  \end{tabular}
  \label{noun_phrases_sentences_example_1} 
\end{table}

\begin{table}[htbp]

  \caption{Example sentences containing interested noun phrases}
  \begin{tabular}{|p{3cm}|p{11cm}|}\hline

    \multirow{5}*{\begin{minipage}{3cm}\textbf{intralobular septa}\end{minipage}} & \footnotesize The lesion density was mostly non-uniform with air bronchogram and thickened interlobular or \textbf{intralobular septa}. \\ \cline{2-2}
    ~ & \footnotesize The lesion may be patchy, nodular, honeycomb, grid or strips, and the lesion density is mostly uneven with the primary presentation of ground glass opacity accompanied by thickening of interlobular or \textbf{intralobular septa}. \\ \cline{2-2}
    ~ & \footnotesize High-resolution CT might show ground glass opacities (figure 5B) in early CT findings (with or without consolidation) followed by \textbf{interlobular septal} and intralobular interstitial thickening with peripheral and lower lobe involvement 93, 94 within the first week of MERS-CoV infection. \\ \cline{2-2}
    ~ & \footnotesize Other findings included intralobular or \textbf{interlobular septal} thickening, and a crazy paving pattern. \\ \cline{2-2}
    ~ & \footnotesize Most of the lesions were distributed along the bronchovascular bundle or the dorsolateral and subpleural part of the lungs and were seen with or without \textbf{interlobular septal} thickening. \\ \hline
    
    \multirow{5}*{\begin{minipage}{3cm}\textbf{trachea}\end{minipage}} & \footnotesize A lateral neck radiograph may show a hazy \textbf{tracheal} air column, with multiple luminal soft tissue irregularities due to pseudomembrane detachment from the soft tissue, but radiographs should be taken only after the patient is stabilized and safe. \\ \cline{2-2}
    ~ & \footnotesize CT imaging demonstrated diffuse edema and narrowing of the glottis, subglottis, and upper \textbf{trachea} in keeping with croup.  \\ \cline{2-2}
    ~ & \footnotesize Tracheal intubation was seen in the \textbf{trachea} and the heart shadow outline was not clear.  \\ \cline{2-2}
    ~ & \footnotesize Black arrows in the \textbf{trachea} indicate cilia and necrosis loss. \\ \cline{2-2}
    ~ & \footnotesize A neck radiograph will show the traditional narrowing of the \textbf{trachea} known as a steeple sign. \\ \hline
    
    \multirow{5}*{\begin{minipage}{3cm}\textbf{bronchiectasis}\end{minipage}} & \footnotesize The timing of c-HRCT scans to diagnose \textbf{bronchiectasis} is important. \\ \cline{2-2}
    ~ & \footnotesize Other c-HRCT signs of \textbf{bronchiectasis} include abnormalities in the surrounding lung may include parenchyma loss, emphysema, scars and nodular foci, 150 a linear array or cluster of cysts, dilated bronchi in the periphery of the lung, and bronchial wall thickening (Box 26.1). \\ \cline{2-2}
    ~ & \footnotesize In particular, CT is much better at identifying the presence and extent of \textbf{bronchiectasis} compared with thoracic radiography. \\ \cline{2-2}
    ~ & \footnotesize (D) Images obtained 5 days later showed partial absorption of the consolidation lesions in the right lower lobe, but fibrosis, \textbf{bronchiectasis}, and vascular thickening occurred. \\ \cline{2-2}
    ~ & \footnotesize In the right upper lobe, there were ground glass densification areas, traction \textbf{bronchiectasis}, interlobular septa thickening and subpleural cystic lesions. \\ \hline

  \end{tabular}
  \label{noun_phrases_sentences_example_2} 
\end{table}



For example, reading the five example sentences containing consolidation, one can judge that consolidation is a typical manifestation of COVID-19. This is in accordance with the conclusion in \citep{consolidation}: ``Consolidation becomes the dominant CT findings as the disease progresses." Similarly, the example sentences of pleural effusion, ground glass opacity, thickening, fibrosis, bronchiectasis, lymphadenopathy show that these abnormalities are closely related with COVID-19. This is consistent with the results reported in the literature:
\begin{itemize}
    \item \textbf{Pleural effusion}: ``In terms of pleural changes, CT showed that six (9.7\%) had pleural effusion." \citep{pleural}
    \item \textbf{Ground glass opacity}: ``The predominant pattern of abnormalities after symptom onset was ground-glass opacity (35/78 [45\%] to 49/79 [62\%] in different periods." \cite{GGO}
    \item \textbf{Thickening}: ``Furthermore, ground-glass opacity was subcategorized into: (1) pure ground-glass opacity; (2) ground-glass opacity with smooth interlobular septal  thickening." \citep{GGO}
    \item \textbf{Fibrosis}: ``In five patients, follow-up CT showed improvement with the appearance of fibrosis and resolution of GGOs.", \citep{fibrosis}
    \item 
     \textbf{Bronchiectasis} and \textbf{lymphadenopathy}: ``The most common patterns seen on chest CT were ground-glass opacity, in addition to ill-defined margins, smooth or irregular interlobular septal thickening, air bronchogram , crazy-paving pattern, and thickening of the adjacent pleura. Less common CT findings were nodules, cystic changes, bronchiolectasis, pleural effusion , and lymphadenopathy."  \citep{air_bronchogram}
\end{itemize}

\section{Conclusions}
In this paper, we develop natural language processing methods to automatically extract unbiased radiological findings of COVID-19. We develop a BERT-based classifier to select sentences that contain COVID-related radiological findings and use bootstrapping to mine hard examples for reducing false positives. Constituent parsing is used to extract noun phrases from the positive sentences and those with top frequencies are selected for medical professionals to further investigate. From the CORD-19 dataset, our method successfully discovers radiological findings that are closely related with COVID-19.


\bibliography{reference}


\end{document}